%% file: hzs.tex
\documentclass[12pt]{iopart}

\begin{document}

\title[Effects of differential mobility on biased diffusion of two species]
{Effects of differential mobility on biased diffusion of two species}
\author{R.S. Hipolito, R.K.P. Zia and B. Schmittmann}

\date{\today }

\begin{abstract}
Using simulations and a simple mean-field theory, we investigate 
jamming transitions in a two-species lattice gas under non-equilibrium
steady-state conditions. The two types of particles diffuse with different 
mobilities on a square lattice, subject to an excluded volume constraint 
and biased in opposite directions. Varying filling fraction, differential 
mobility, and drive, we map out the phase diagram, identifying first order and 
continuous transitions between a free-flowing disordered and a spatially 
inhomogeneous jammed phase. Ordered structures are observed to drift, with
a characteristic velocity, in the direction of the more mobile species. 
\end{abstract}

\address{Center for Stochastic Processes in Science and Engineering, Physics Department \\ 
	Virginia Polytechnic Institute and State University, \\ 
	Blacksburg, VA 24061-0435, USA}

\jl{1}



\section{Introduction}

Since their introduction, driven lattice gases have attracted much attention
as some of the simplest systems for exploring non-equilibrium steady states %
\cite{KLS,DL17}. Motivated by the physics of fast ionic conductors \cite{FIC}%
, the earliest models consisted of a single species of ``charged''
particles, diffusing on a periodic lattice, subject to the effects of both a
thermal bath and an external (uniform, DC) ``electric'' field. Driven by the
field, the system settles into a state with non-trivial current. With no
inter-particle interactions other than an excluded volume constraint, the 
{\em stationary state} distribution is trivial \cite{Spitzer}, and
interesting behavior is displayed only in {\em time-dependent} quantities.
However, once interactions are introduced, then a variety of unexpected
phenomena arise even in the steady state, such as long-range correlations at 
{\em all }temperatures \cite{LRC}.

A natural generalization is to study systems with two species, motivated by,
e.g., fast ionic conductors with several mobile species \cite{FIC}. Such
systems have been used to model water droplets in microemulsions with
distinct charges \cite{water-in-oil}, gel electrophoresis \cite%
{gel-electro,Mukamel-gel}, vacancy mediated diffusion in binary alloys \cite%
{VMD} and traffic flow \cite{traffic}. Focusing on systems with two species
carrying equal but opposite ``charge'', we discovered remarkably complex
steady states, even for particles subjected only to the excluded volume
constraint \cite{HSZ,BSZ}. For simplicity, the early studies employed only
square lattices of $L\times L$, filled with equal numbers of ``positive''
and ``negative'' particles. The control parameters in this simple system are
just $E$, the strength of the drive, and $\bar{m}$, the overall particle
density (regardless of charge). By varying these parameters, we found that
this model displays a phase transition \cite{HSZ}, from a disordered state
with homogeneous densities at small $E$ or $\bar{m}$, to an ordered state
with inhomogeneous densities. Moreover, this line in the $E$-$\bar{m}$ plane
consists of a section with discontinuous transitions and another with
continuous ones, which may be regarded as analogues of first and second
order transitions in equilibrium systems. To understand these phenomena, it
is sufficient to adopt a mean-field, continuum approach, based on
hydrodynamic like equations for the two conserved densities. A linear
stability analysis around the homogeneous state allows us to predict the 
{\em existence} of transitions. Fortunately, we were able to find {\em %
analytic }solutions for both, homogeneous and inhomogeneous steady states %
\cite{VZS}, so that the presence of {\em both types} of transitions can be
predicted. Subsequently, such models have been generalized to include
rectangular lattices \cite{BSZ}, non-neutral \cite{LZ} or nearly filled \cite%
{ST} systems, and ``charge exchange'' dynamics \cite{KSZ97} in
one-dimensional \cite{Bonn} and quasi-one-dimensional \cite{KSZ99, MSZ}
cases. A dazzling array of surprising phenomena was discovered.

Here, we investigate a further generalization, namely, having species with
different mobilities. Indeed, unless there are some underlying symmetry
constraints, there is no reason to expect two different species of particles
in typical physical systems to have identical mobilities. Using both Monte
Carlo techniques and mean-field analyses, we find that, though the phase
diagram appears to suffer little change, novel features arise in all regions
of parameter space, even at not-so-subtle levels. In the next section, we
present some details of how a differential mobility is implemented in
simulations, followed by a discussion of our results. Section 3 is devoted
to the continuum, mean-field theory and its predictions. We close with a
summary of our findings and discuss some possible avenues for future
investigations.

\section{Model Specification and Simulation Results}

Following the first studies of biased diffusion of two species, square
lattices of size $L\times L\equiv N$ are used. In most of our simulations, $%
L=$ $40$. We focus exclusively on neutral systems (i.e., $N_{+}=N_{-}$) at
various densities $\bar{m}\equiv \left( N_{+}+N_{-}\right) /N$. A
configuration of the system is specified by the set $\left\{ \sigma
_{x,y}\right\} $, where $\sigma _{x,y}$ assumes the values $0,\pm 1$ if the
site $\left( x,y\right) $ is empty or occupied by a positive ($+$) or
negative ($-$) particle. Clearly, $\bar{m}=\sum_{x,y}\sigma _{x,y}^{2}/N$.
The diffusive nature of the the particles is modeled by allowing them to
jump only to nearest-neighbor empty sites. After a particle and one of its
neighboring site are selected, a particle-hole pair is always exchanged,
unless it results in a particle jumping against the external ``electric''
field. Chosen to point in the positive $y$ axis, the field is parametrized
by $\vec{E}=E\hat{y}$, so that its sole effect is to {\em lower} the
probability of particle jumps against $\vec{E}$ to $e^{-E}$.

In contrast to previous studies, the two species are endowed with {\em %
different mobilities}, modeling say, trucks and cars on a road. Using the
self-evident notation $\mu _{\pm }$ for the mobilities, we arbitrarily
choose $\mu _{+}\geq \mu _{-}$ and then focus on the ratio $\mu \equiv \mu
_{+}/\mu _{-}$. To implement this ratio, we keep a list of the particles and
their locations. In one Monte Carlo step (MCS), $\bar{m}N$ particles are
chosen from the positive/negative list randomly with the frequency ratio of $%
\mu $. In particular, by focusing on integer values of $\mu $, we simply
make $\mu $ attempts to move a positive particle for every attempt to move a
negative one. By monitoring a few quantities, we found that there is little
difference between sequential and random updating in species space,
especially since the particle locations were randomly assigned at the
beginning. After $\bar{m}N$ attempts, we increment the time by 1 MCS. Below,
we will see that it is often convenient to use the ``renormalized'' MCS (1
RMCS being defined as $\mu +1$ MCS), during which every slow particle would
have had, on the average, one chance to move. Typically, 20K MCS or more are
discarded before measurements are taken. In this study, we probe $\mu
=1,3,10,30,100,$ and $300$, though most of the data are collected for $\mu
=30$, being a compromise between exploring large differentials and dealing
with finite computation times.

\subsection{Continuous and discontinuous phase transitions}

Similar to the $\mu =1$ system, ours displays a transition from a disordered
homogeneous to an ordered inhomogeneous state, the latter showing a
marcoscopic cluster with opposing particles locked in a ``jam.'' To
differentiate these phases, we monitor the first few {\em mass }structure
factors, defined as 
\[
S_{n}\equiv \left\langle \left| \tilde{m}_{n}\right| ^{2}\right\rangle , 
\]
where 
\[
\tilde{m}_{n}\equiv L^{-2}\sum_{x,y}\sigma _{x,y}^{2}\exp \left[ 2\pi iny/L%
\right] 
\]
is the Fourier tranform of the mass density profile across $y$. Here, $%
\left\langle \cdot \right\rangle $ denotes the average over 800
configurations, separated from one another by 100 MCS in a typical run of
80K MCS. We also measure the fluctuations in the $S$'s by monitoring the
variances: $\left\langle \left[ \left| \tilde{m}_{n}\right| ^{2}-S_{n}\right]
^{2}\right\rangle $. Both continuous and discontinuous transitions are
observed, as illustrated by a phase diagram in the $E$-$\bar{m}$ plane for
the $\mu =30$ case (Fig.~1). The insets illustrate the nature of the
transitions, in two typical parameter domains. 

\begin{figure}[tbp]
\input{epsf}
\par
\begin{center}
\begin{minipage}{.5\textwidth}
    \vspace{-1.cm}
  \epsfxsize = \textwidth \epsfysize = .9\textwidth \hfill
  \epsfbox{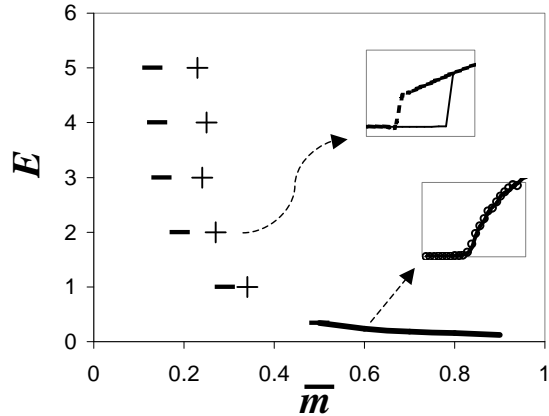}
    \vspace{-1.cm}
\end{minipage}
\end{center}
\caption{Phase diagram for mobility ratio of 30, with the system being homogeneous
and ``jammed'' in the lower-left and upper-right regions, respectively. The
solid line marks continuous transitions. For discontinuous transitions, a
pair of symbols ($+/-$) are used to denote the jumps in the order parameter
when the filling fraction ($\bar{m}$) is increased/decreased. The
absence/presence of hysteresis loops is illustrated by the insets, each
containing data for sweeps in {\em both }directions (cf. Fig.~2 for further
explanations).}
\end{figure}

To provide further detail on how these lines are determined, we expand the
insets of Fig.~1 into Fig.~2a,b. Here, we plot the variations in $S_{1}$ as
the system is swept {\em back and forth} across the transitions. For the
continuous case (Fig.~2a), we show the result from a sweep in $E\in \left[
0,0.5\right] $ while $\bar{m}$ is held at $0.6$. This direction of
traversing the transition line is chosen because this line is almost
parallel to the $\bar{m}$-axis in this region. Note that there are two sets
of data points ($\times /-$), representing the up/down sweeps from a single
long run. This run is started with $E=0$ in a random configuration at just
over half filling ($\bar{m}=0.6$). Then, the first 20K MCS are discarded,
data gathered for the next 80K MCS, $E$ raised by $0.02$, 20K discarded and
so on, until $E=1.0$. Thereafter, the process is reversed, decreasing $E$
until $E=0$. The figure shows clearly that the transition is continuous with
no sign of hysteresis. At the crude level of our investigations, the
critical field is then identified through either the largest $\partial
S_{1}/\partial E$ or the peak in the variance of $S_{1}$. The transition
line is assembled by repeating such runs for $\bar{m}\in \left[ 0.5,0.95%
\right] $ in steps of $0.05$. For the discontinuous cases, which are more
easily probed by sweeping in $\bar{m}$, we observe the typical hysteresis
loops. The inset in Fig.~1 and Fig.~2b show the case of $E=2$. Again, the
data represent a single long run, starting with $\bar{m}=0.1$ in a random
configuration, discarding the first 20K MCS, gathering data for the next 80K
MCS as above, raising $\bar{m}$ by $0.01$ and so on. These data points are
represented by crosses ($\times $). After a run at $\bar{m}=0.4$, we
continue this process with decreasing $\bar{m}$'s until $\bar{m}=0.1$. This
set of data is shown as minus signs ($-$). A hysteresis loop is clearly
displayed, and the densities at the jumps are recorded to construct the
points plotted in the main figure of Fig.~1. For comparison with the
previous studies (i.e., the $\mu =1$ case), we also show the results of such
runs as solid lines in Fig.~2a,b. As we see, there is little difference
between the $\mu =30$ and $\mu =1$ results, leading us to conclude that,
within the accuracy and statistics of our study, there is no significant
dependence of the phase diagram on $\mu$.

\begin{figure}[tbp]
\input{epsf}
\par
\begin{center}
\vspace{-7.cm}
\begin{minipage}{.9\textwidth}
  \epsfxsize = \textwidth \epsfysize = .9\textwidth \hfill
  \epsfbox{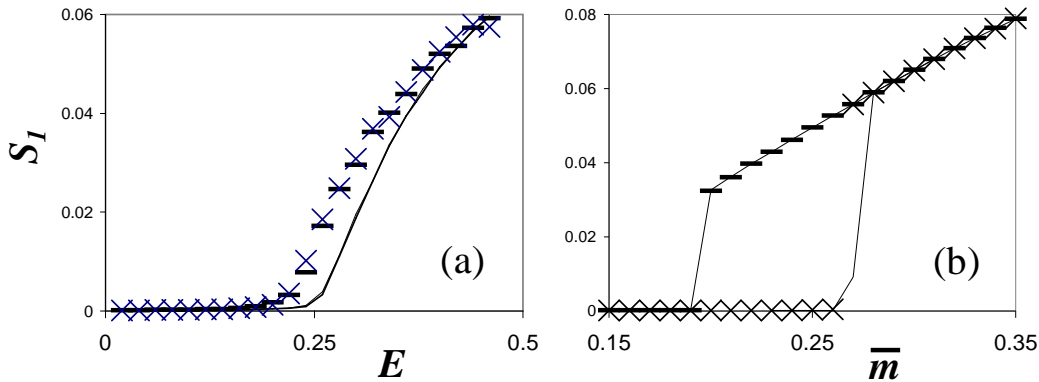}
    \vspace{-1.cm}
\end{minipage}
\end{center}
\caption{Plots of $S_{1}$ as the system is swept across the transitions. (a) $E$ is
increased ($\times $) and then decreased ($-$), with $\bar{m}=0.6$ and $\mu
=30$. (b) $\bar{m}$ is increased ($\times $) and then decreased ($-$), with $%
E=2$ and $\mu =30$. For comparison, the $\mu =1$ cases are shown as solid
lines in both panels. With our statistics and accuracy, the effects of
mobility differential are relatively minor.}
\end{figure}

Before turning to the properties of the inhomogeneous state, let us remark
that, of course, there {\em are} non-trivial effects due to $\mu >1$, even
in the disordered phase. Since one species is more mobile, there must be a
difference between the average velocities, which in turn leads to a non-zero 
{\em mass} current even though $N_{+}=N_{-}$. To verify this expectation, we
measure the two currents, $J_{\pm }$, by simply keeping track of the number
of $+/-$ particles which move up/down. Since $N_{+}=N_{-}$ in our samples,
the ratio of the average velocities is just $v_{+}/v_{-}=J_{+}/J_{-}$. With
limited observations, our naive expectation: 
\[
J_{+}/J_{-}=\mu 
\]%
turns out to be satisfied quite well. As the data in Table 1 show, we have
investigated only a few sample points, in both ``arms'' of the disordered
region (small $E$ moderate $\bar{m}$, and vice versa).

\begin{center}
\begin{tabular}{|r|r|r|r|r|r|}
\hline
$E$ & $\bar{m}$ & $\mu $ & $J_{+}/J_{-}$ & $J_{+}$ & $J_{-}$ \\ \hline
$0.1$ & $0.5$ & $3$ & $5.00$ & $0.22$ & $0.04$ \\ \hline
$0.2$ & $0.5$ & $3$ & $2.47$ & $0.39$ & $0.16$ \\ \hline
$0.1$ & $0.5$ & $30$ & $34.8$ & $2.35$ & $0.07$ \\ \hline
$0.2$ & $0.5$ & $30$ & $31.9$ & $4.44$ & $0.14$ \\ \hline
$2.0$ & $0.1$ & $3$ & $3.01$ & $1.13$ & $0.38$ \\ \hline
$2.0$ & $0.2$ & $3$ & $2.91$ & $1.81$ & $0.62$ \\ \hline
$2.0$ & $0.1$ & $30$ & $31.0$ & $11.1$ & $0.36$ \\ \hline
$2.0$ & $0.2$ & $30$ & $28.9$ & $18.4$ & $0.64$ \\ \hline
\end{tabular}

\vspace{0.5cm}
Table 1. Currents and their ratios for various regions in the homogeneous
phase.

\medskip
\end{center}

Naturally, we expect other interesting properties associated with the
disordered state; yet, these are generally quite subtle. For example, even
in the $\mu =1$ case, there are singular structure factors and long-range
correlations \cite{KSZ97b}. For more prominent and fascinating features, we
turn to the inhomogeneous state.

\subsection{Drifting structures in the ordered state}

Here, the density of particles is high enough so that they ``jam'' into a
macroscopic cluster. Consequently, the translational symmetry (in $y$) is
spontaneously broken. In neutral systems with $\mu =1$, the cluster performs
an {\em unbiased} random walk, and, due to the symmetry under charge
``conjugation'' ($+\Leftrightarrow -$), its charge profile is purely
antisymmetric. On the other hand, for charged systems (i.e., $N_{+}\neq
N_{-} $) this cluster drifts ``{\em backwards}'' (opposite to the average
motion of the majority species)! Though counter-intuitive at first glance,
this behavior can be understood \cite{LZ}. Simulating $\mu >1$ systems here,
we find the cluster to drift ``{\em forwards},'' i.e., in the direction of
the more mobile species. Though this behavior may seem less surprising, it
is easy to advance a (wrong) heuristic argument to arrive at the
opposite conclusion!

\begin{figure}[tbp]
\input{epsf}
\par
\begin{center}
\vspace{-1.cm}
\begin{minipage}{.5\textwidth}
  \epsfxsize = \textwidth \epsfysize = .9\textwidth \hfill
  \epsfbox{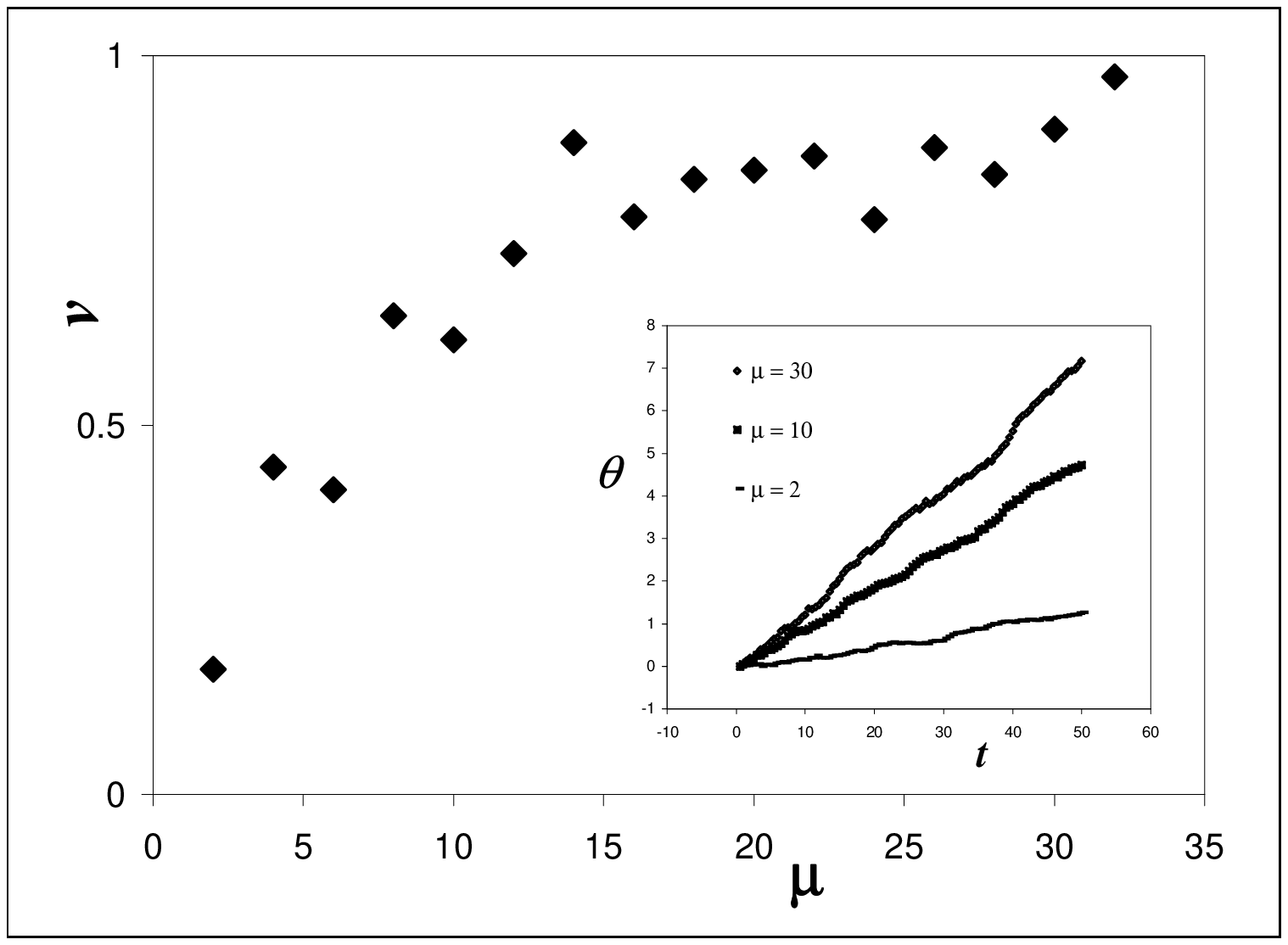}
    \vspace{-1.cm}
\end{minipage}
\end{center}
\caption{Drift speed as a function of $\mu $. The inset shows the ``center of mass''
of the macroscopic cluster, as a function of time, for three values of $\mu$.}
\end{figure}

To characterize the drifting cluster more systematically, we carry out the
following analysis. Since the runs are necessarily long, we focus only on
neutral lattices with $\bar{m}=0.36$, driven with $E=1$. These parameters
put us well in the ordered phase. To study the $\mu $-dependence of the
drift speed, we perform simulations with $\mu \in \left[ 2,32\right] $ in
steps of $2$. For comparison with earlier studies, we also collect data for $%
\mu =1.$ Using a random configuration at the start, the usual first 20K MCS
are discarded. To be sure that a macroscopic cluster is present, we first
test the magnitude $\left| \tilde{m}_{1}\right| $ and proceed only when it
is larger than $0.25$ (c.f., $\left| \tilde{m}_{1}\right| =1\left/ \left[
L\sin (\pi /L)\right] \right. \simeq 0.32$ for a fully ordered state).
Assuming the center of mass (CM) of the cluster is well characterized by the
CM of the entire configuration, we measure the latter at intervals of 500
RMCS. Here, the CM of a particular configuration is defined simply by the
phase in $\tilde{m}_{1}=|\tilde{m}_{1}|e^{i\theta }$. Also, we find it more
convenient to use the RMCS as a unit, since the time scale of the system as
a whole is really set by the slow-moving particles. In the inset of Fig. 3,
we show how $\theta $ depends on time for three $\mu $'s. Clearly, the
dominant behavior is a steady drift. Fitting these data to $v_{\theta
}t+const$, we extract the drift speed $v_{\theta }$. Translating $v_{\theta
} $ into a velocity, $v$, in units of lattice spacing per 1000 RMCS, we plot 
$v\left( \mu \right) $ in Fig. 3. Until we have some understanding of this
quantity, we refrain from fitting it to a particular form. Instead, we just
note that, interestingly, $v$ seems to saturate at unity. Finally, we have
computed the standard deviations from the linear fits naively and found that
these are essentially independent of $\mu $. Of course, since we believe the
CMs to be performing (biased) random walks, we should expect these
deviations to grow as $t^{1/2}$, from which diffusion coefficients can be
extracted. Postponing such a detailed study to a later publication, what we
may infer from our ``naive'' computation is just that clusters associated
with different $\mu $'s execute random walks with the {\em same} diffusion
constant but {\em different} biases.

Next, we consider the mass and charge profiles in the jammed state. For each
configuration we measure both $\theta $ and 
\[
m\left( y\right) \equiv \sum_{y}\sigma _{x,y}^{2} \quad and \quad 
q\left( y\right) \equiv \sum_{y}\sigma _{x,y}\,\,. 
\]%
Then we displace the $y$-dependent quantities according to $y\rightarrow
u\equiv y-\theta L/2\pi +L/2$, so that the CM is located at $L/2=20$ for
convenience. In this co-moving frame, we can now average over the run to
produce the profiles: $\left\langle m\left( u\right) \right\rangle $ and $%
\left\langle q\left( u\right) \right\rangle $. Plotted in Fig. 4 are these
profiles for the parameter set $\left( E,\bar{m},\mu \right) =\left(
1,0.36,30\right) $, and, for comparison with previous results, those for the 
$\mu =1$ case as well. Even though the change in the mass profile appears
almost imperceptible, there are actually some significant differences. For
example, though the densities within (outside) the cluster differ by only
about $+0.003\left( -0.003\right) $, this discrepancy translates into a {\em %
factor of two} for the densities far from the ``jam.'' Until we have much
better statistics, it is unclear if we can attach much meaning to these
subtle differences. The changes in the charge profile are more discernable.
We may interpret these as an enhancement of the more mobile species' ability
to penetrate into the block of the less mobile one.

\begin{figure}[tbp]
\input{epsf}
\par
\begin{center}
\vspace{1.cm}
\begin{minipage}{.5\textwidth}
  \epsfxsize = \textwidth \epsfysize = .9\textwidth \hfill
  \epsfbox{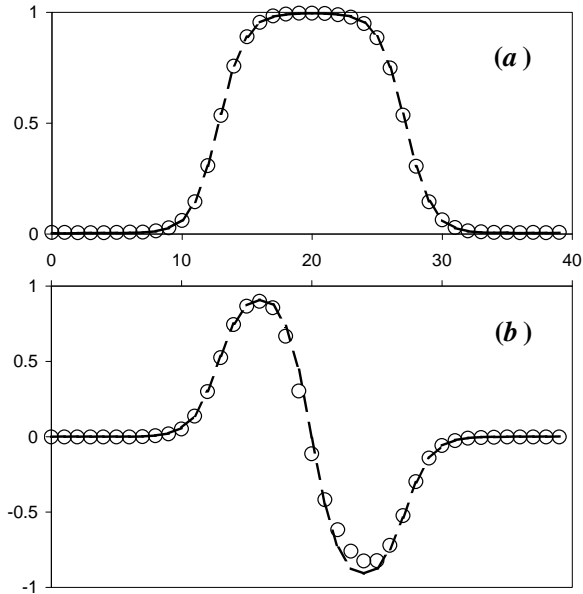}
    \vspace{-1.cm}
\end{minipage}
\end{center}
\caption{Mass (a) and charge (b) profiles in the ordered state. The dashed line and
the open circles correspond to the $\mu =1$ and $\mu =30$ cases,
respectively. Note that the charge profile for the latter is not purely
antisymmetric about the CM ($L/2=20$).}
\end{figure}

\section{Mean-Field Theory and Comparisons}

To develop a first-step understanding of the large scale phenomena shown
above, we exploit the same mean field approach which proved quite successful
in the $\mu =1$ case \cite{HSZ,VZS}. In this approach, hydrodynamic like
equations of motion are postulated for the densities of the two species, $%
\rho _{\pm }\left( \vec{x},t\right) \equiv \left\langle \sigma ^{2}\left( 
\vec{x},t\right) \pm \sigma \left( \vec{x},t\right) \right\rangle /2$ : 
\[
\frac{\partial \rho _{\pm }}{\partial t}=-\vec{\nabla}\cdot \vec{J}_{\pm } 
\]
where $\vec{J}_{\pm }$ are the associated (density dependent) currents \cite%
{HSZ}. Here, we also treat the spatial co-ordinates as continuous variables,
in general $d$ dimensions: $\vec{x}=\left( x_{1},...,x_{d}\equiv y\right) $.
To model differential mobility, we simply multiply the currents by $\mu
_{\pm }$. Absorbing the average mobility into the time scale by defining 
\[
\mu _{\pm }\equiv \left( 1\pm \delta \right) \,\,, 
\]
the continuum equations read 
\begin{equation}
\frac{\partial \rho _{\pm }}{\partial t}=\left( 1\pm \delta \right) \vec{%
\nabla}\cdot \left[ \phi \vec{\nabla}\rho _{\pm }-\rho _{\pm }\vec{\nabla}%
\phi \mp \rho _{\pm }\phi {\cal E}{\bf \hat{y}}\right] \quad ,  \label{EoM}
\end{equation}
where 
\[
\phi \equiv 1-\rho _{+}-\rho _{-} 
\]
is the hole density and ${\cal E}$ represents the coarse-grained electric
field (${\cal E}=2\tanh \left( E/2\right) $ at the most naive level). Of
course, these equations can be re-expressed in terms of $\phi $ and the
charge density 
\[
\psi \equiv \rho _{+}-\rho _{-} 
\]
resulting in a form 
\begin{equation}
\frac{\partial }{\partial t}\left( 
\begin{array}{c}
\phi \\ 
\psi%
\end{array}
\right) =\left[ 
\begin{array}{cc}
1 & -\delta \\ 
-\delta & 1%
\end{array}
\right] \left( 
\begin{array}{c}
\nabla \cdot \left[ \nabla \phi +\phi \psi {\cal E}{\bf \hat{y}}\right] \\ 
\nabla \cdot \left[ \phi \nabla \psi -\psi \nabla \phi -\phi (1-\phi ){\cal E%
}{\bf \hat{y}}\right]%
\end{array}
\right)  \label{phi-psi EoM}
\end{equation}
which easily reduces to the one studied previously \cite{HSZ,VZS}.

To find solutions to these equations, we must supply boundary conditions and
constraints. For a hypercubic system of volume $L^{d}$, they are 
\[
\rho _{\pm }\left( \vec{x},t\right) =\rho _{\pm }\left( \vec{x}+L\hat{x}%
_{i},t\right) \,\, , i=1,2,...,d 
\]
and 
\[
\int \frac{d\vec{x}}{L^{d}}\phi \left( \vec{x},t\right) =1-\bar{m},\quad
\int \frac{d\vec{x}}{L^{d}}\psi \left( \vec{x},t\right) =0. 
\]
We will discuss mainly steady states, i.e., density profiles which are time
independent in a {\em co-moving }frame. From Monte Carlo results as well as
previous experience with this model, we expect these profile to obey
translational invariance in the transverse directions in the whole parameter
space. On the other hand, spontaneous breaking of translational invariance
in $y$ is manifest for the jammed state. Due to the drift, we will denote
these profiles by $\phi \left( u\right) $ and $\psi \left( u\right) $, with $%
u=y-vt.$

\subsection{Homogeneous states and linear stability analysis}

Let us consider first the disordered phase, where the steady-state densities
are homogeneous in both space and time, i.e., 
\[
\phi \left( \vec{x},t\right) =1-\bar{m},\quad \psi \left( \vec{x},t\right)
=0. 
\]%
Note that, unlike the $\mu =1$ case, the mass current is no longer zero. The
(average), denoted by $C\hat{y}$, is given through 
\[
C\equiv \hat{y}\cdot \left( \vec{J}_{+}+\vec{J}_{-}\right) =\mu _{+}{\cal E}%
(1-\bar{m})\bar{m}/2-\mu _{-}{\cal E}(1-\bar{m})\bar{m}/2=\delta \bar{m}(1-%
\bar{m}){\cal E}. 
\]%
With our normalization of time scales, the charge current is unchanged: 
\[
J\equiv \hat{y}\cdot \left( \vec{J}_{+}-\vec{J}_{-}\right) =\bar{m}(1-\bar{m}%
){\cal E}. 
\]%
Some care is needed when direct comparisons of absolute currents with Monte
Carlo data are made, since MCS and RMCS differ by a factor of $\mu +1$. One
test of this theory which is independent of such details is the ratio of
current magnitudes 
\[
J_{+}/J_{-}=\mu . 
\]%
As noted above, this prediction agrees well with our limited observations
(apart from the first set, in which both currents are very small and noisy).
Of course, our crude data should be regarded as merely the first step in a
more refined study. That they are consistent with $J_{+}/J_{-}=\mu $ just
confirms that this naive way of introducing mobility differentials is a
reasonable starting point. Noting that the agreement is somewhat worse for
high densities suggests the presence of subtle particle-particle
correlations, which can only be uncovered by better statistics.

Beyond these trivial results, a linear stability analysis around this state
leads us to the matrix 
\[
M=\left[ 
\begin{array}{cc}
1 & -\delta \\ 
-\delta & 1%
\end{array}%
\right] \left[ 
\begin{array}{cc}
-k^{2} & ik_{y}{\cal E}(1-\bar{m}) \\ 
ik_{y}{\cal E}(2\bar{m}-1) & -k^{2}(1-\bar{m})%
\end{array}%
\right] 
\]%
which is purposely written in a form to display the effects of $\delta >0$.
In contrast to the $\delta =0$ case, the eigenvalues ($\lambda _{1,2}$) of
this $M$ are generically complex. Thus, we should examine their real
parts to identify the stability limit. However, since $M$ is
real, it can be expressed as $\left| \lambda _{1}\right| ^{2}%
\mathop{\rm Re}%
\lambda _{2}/%
\mathop{\rm Re}%
\lambda _{1}$ (or $1\Leftrightarrow 2$), so that $\det M=0$ is still
valid for identifying the stability limit. Since $\det M$ is modified
by a simple factor of $\left( 1-\delta ^{2}\right) $, we conclude that
differential mobility has {\em no effect} on this line! In other words, the
most unstable perturbation is still given by the lowest longitudinal mode: 
\[
k_{y}=2\pi /L 
\]%
and the instability occurs at \cite{HSZ} 
\[
\left( {\cal E}L/2\pi \right) ^{2}(2\bar{m}-1)=1. 
\]%
Of course we cannot expect good quantitative comparisons with the phase
boundaries from Monte Carlo simulations. For example, in the Ising model,
mean-field analysis overestimates the critical temperature by nearly a
factor of 2. Nevertheless, we are encouraged by one aspect of this
prediction, namely, that differential mobility has no effect on the onset of
the instability. This seems to be consistent with the observed phase
boundaries being quite insensitive to $\mu $.

For completeness, we give the explicit expressions for the eigenvalues: 
\begin{eqnarray}
\lambda _{1,2}=-k^{2}(1-\bar{\rho})-\delta ik_{y}{\cal E}\bar{\rho} 
\pm \sqrt{R}
\end{eqnarray}
where
\begin{eqnarray}
R = k^{4}\left[ 
\bar{\rho}^{2}+\delta ^{2}(1-\bar{m})\right] &-& (k_{y}{\cal E})^{2}\left[
(1-\delta ^{2})(2\bar{m}-1)(1-\bar{m})+\delta ^{2}\bar{\rho}^{2}\right] 
\nonumber \\
 &+& 2\delta ik^{2}k_{y}{\cal E}\bar{\rho}(1-\bar{\rho})
\end{eqnarray}
and $\bar{\rho}\equiv \bar{m}/2$. Their complex nature reflects drifts
associated with the eigen-perturbations. Given that the species are mobile
in different ways, such drifts are to be expected. In principle, it is
possible to measure the decay and drift of small perturbations in
simulations so that comparisons with these predictions can be made. In
practice however, such measurements would require better statistics and
longer runs than those in this study.

\subsection{Density profiles in inhomogeneous states}

Turning to the more interesting ordered states, the most immediate question
concerns the steady-state profiles. As observed in simulations and expected
from symmetry considerations, these profiles are homogeneous in $x$ but
depend on $y$. In analogy to the $\mu =1$ case, we seek non-trivial
functions $\phi ^{\ast }\left( y\right) $ and $\psi ^{\ast }\left( y\right) $
which satisfy Eqn. (\ref{phi-psi EoM}) in a steady state. However, it is
clear from Eqn. (\ref{EoM}) that if non-trivial functions lead to zeroes on
the right hand side with $\mu =1$, then they also lead to zeroes for any $%
\mu $! The conclusion is that our mean-field theory admits inhomogeneous
profiles which are (i) stationary 
(i.e., do not drift) and (ii) identical to those for $\mu =1$.
Of course, we may argue that this obvious discrepancy,
when comparing with data, is a ``small'' effect in absolute terms. However,
our equations are non-linear and the existence of additional, drifting
solutions cannot be ruled out. Unfortunately, we are unable so far, either
to find drifting solutions (analytically or numerically), or to prove that
our equations admit no such solutions. Nevertheless, for completeness, let
us present the analysis which simplifies the problem to a single second
order, non-linear, ordinary differential equation.

Following the techniques in \cite{LZ}, we assume the forms 
\[
\left( 
\begin{array}{c}
\phi ^{\ast }\left( \vec{x},t\right) \\ 
\psi ^{\ast }\left( \vec{x},t\right)%
\end{array}%
\right) =\left( 
\begin{array}{c}
\phi ^{\ast }\left( u\right) \\ 
\psi ^{\ast }\left( u\right)%
\end{array}%
\right) 
\]%
where 
\[
u\equiv y-vt, 
\]%
with $v$ being the drift velocity of the macroscopic cluster. Inserting
these into Eqn. (\ref{phi-psi EoM}), we obtain 
\[
-v\frac{d}{du}\left( 
\begin{array}{c}
\phi ^{\ast } \\ 
\psi ^{\ast }%
\end{array}%
\right) =\frac{d}{du}\left[ 
\begin{array}{cc}
1 & -\delta \\ 
-\delta & 1%
\end{array}%
\right] \left( 
\begin{array}{c}
\left[ \left( d\phi ^{\ast }/du\right) +\phi ^{\ast }\psi ^{\ast }{\cal E}%
\right] \\ 
\left[ \phi ^{\ast }\left( d\psi ^{\ast }/du\right) -\psi ^{\ast }\left(
d\phi ^{\ast }/du\right) -\phi ^{\ast }(1-\phi ^{\ast }){\cal E}\right]%
\end{array}%
\right) , 
\]%
which has the form of a vanishing total derivative. So, the first integrals
are just constants which are identified with the two currents. 
As presented above, we denote the mass current $C$ and
the charge-current by $J$.
Recognizing that the {\em hole} current must be $-C$, we obtain 
\begin{eqnarray}
\left( 
\begin{array}{c}
C \\ 
-J%
\end{array}%
\right) &=& v\left( 
\begin{array}{c}
\phi ^{\ast } \\ 
\psi ^{\ast }%
\end{array}%
\right) 
 \\
&+& \left[ 
\begin{array}{cc}
1 & -\delta \\ 
-\delta & 1%
\end{array}%
\right] \left( 
\begin{array}{c}
\left[ \left( d\phi ^{\ast }/du\right) +\phi ^{\ast }\psi ^{\ast }{\cal E}%
\right] \\ 
\left[ \phi ^{\ast }\left( d\psi ^{\ast }/du\right) -\psi ^{\ast }\left(
d\phi ^{\ast }/du\right) -\phi ^{\ast }(1-\phi ^{\ast }){\cal E}\right]%
\end{array}%
\right) . \nonumber 
\end{eqnarray}
To simplify these equations, we change variables, as in the $\delta =0$
case, to 
\[
\chi \equiv 1/\phi ^{\ast }\quad and \quad \omega \equiv \psi ^{\ast
}/\phi ^{\ast }. 
\]%
Then, we invert the matrix so that the first derivatives are decoupled: 
\[
\left( 
\begin{array}{c}
-d\chi /du+\omega {\cal E} \\ 
d\omega /du-(\chi -1){\cal E}%
\end{array}%
\right) =\frac{-\chi }{1-\delta ^{2}}\left[ 
\begin{array}{cc}
1 & \delta \\ 
\delta & 1%
\end{array}%
\right] \left( 
\begin{array}{c}
-C\chi +v \\ 
J\chi +v\omega%
\end{array}%
\right) . 
\]%
Being first order differential equations, their full solutions will require
two more unknown integration constants. Together with $C$ and $J$, there are
four unknowns to be fixed, by the four constraint equations: 
\[
\chi \left( 0\right) =\chi \left( L\right) \quad and \quad \omega
\left( 0\right) =\omega \left( L\right) 
\]%
from periodic boundary conditions, as well as 
\[
\int_{0}^{L}\frac{1}{\chi }du=\int \phi ^{\ast }=L\left( 1-\bar{m}\right)
\quad and \quad \int \frac{\omega }{\chi }=\int \psi ^{\ast }=0 
\]%
from the conservation laws. Further simplifications occur when we define $%
C,J,v$ with appropriate factors of ${\cal E}$, $\delta $, and $1-\delta
^{2}: $%
\begin{eqnarray*}
\frac{C}{1-\delta ^{2}} &\equiv &\delta {\cal E}\tilde{C} \\
\frac{J}{1-\delta ^{2}} &\equiv &{\cal E}\tilde{J} \\
\frac{v}{1-\delta ^{2}} &\equiv &\delta {\cal E}\tilde{v}
\end{eqnarray*}%
so as to exploit the scaling of $u$ to the variable $u{\cal E}$, and the
symmetry of the system under $\delta \Leftrightarrow -\delta $ (with $\tilde{%
C},\tilde{J},\tilde{v}$ even in $\delta $). Finally, denoting 
$d/d(u{\cal E})$ by prime ($^{\prime }$), we arrive at 
\begin{eqnarray}
\chi ^{\prime } &=&\omega +\delta \left( \tilde{J}-\tilde{C}\right) \chi
^{2}+\delta \tilde{v}\chi \left( 1+\delta \omega \right)  \label{DEchi} \\
\omega ^{\prime } &=&(\chi -1)-\left( \tilde{J}-\delta ^{2}\tilde{C}\right)
\chi ^{2}-\delta \tilde{v}\chi \left( \delta +\omega \right) .
\label{DEomega}
\end{eqnarray}%
One advantage of this form lies in that, in the limit $\delta \rightarrow 0$%
, it easily reduces to the previous case: $\chi ^{\prime \prime }=(\chi -1)-%
\tilde{J}\chi ^{2}$. The other is the clear separation of $\chi ,\omega $
into even/odd functions of $\delta $. Note that, under this ``parity''
operation, $u{\cal E}$ is odd.

As in the $\delta =0$ case, the first equation can be trivially solved for $%
\omega $ and the resultant inserted into the second, leading us to a single
(though quite complicated) equation for $\chi .$ As pointed out above, there
exists at least {\em one} solution for any $\delta $, corresponding to a
stationary ($v=0$) state and involving $\tilde{C}=\tilde{J}$ (i.e., $%
J=\delta C$).

Although we have not found non-trivial drifting solutions analytically, we
are able to test the consistency of these equations against the measured
profiles and $v$. For example, we can begin with Eqn.(\ref{DEchi}) and
integrate over $u$ to get relationships between $v$ and averages of the
profiles. Indeed, there are infinitely many similar such relations, obtained
from first multiplying this equation by functions of $\chi $ (or $\phi
^{\ast }$) before integration. To minimize possible large errors, we choose
to consider two equations so that the averages involve no higher power of $%
\chi $ than unity. Specifically, we {\em divide} Eqn.(\ref{DEchi}) by $\chi $
and $\chi ^{2}$ before integration. The results are 
\begin{eqnarray*}
0 &=&\left\langle \psi \phi \right\rangle +\delta \left( \tilde{J}-\tilde{C}%
\right) +\delta \tilde{v}\left( 1-\bar{m}\right) \\
0 &=&\delta \left( \tilde{J}-\tilde{C}\right) \left\langle \chi
\right\rangle +\delta \tilde{v}\left( 1+\delta \left\langle \omega
\right\rangle \right)
\end{eqnarray*}%
where the average is defined via 
\[
\left\langle \bullet \right\rangle \equiv \frac{1}{L}
\int_{0}^{L} \bullet \, du 
\]%
Eliminating the unknown currents, we arrive at 
\[
\frac{v/{\cal E}}{1-\delta ^{2}}=\frac{\left\langle \psi \phi \right\rangle
\left\langle \chi \right\rangle }{1-\left\langle \phi \right\rangle
\left\langle \chi \right\rangle +\delta \left\langle \psi ^{\ast }\chi
\right\rangle } 
\]%
To compare with data, we retrace the rescaling of time and re-introduce the
two mobilities, so that 
\[
v=\frac{2\mu _{+}\mu _{-}\left\langle \chi \right\rangle \left\langle \psi
^{\ast }\phi ^{\ast }\right\rangle {\cal E}}{\left( \mu _{+}+\mu _{-}\right) %
\left[ 1-\left( 1-\bar{m}\right) \left\langle \chi \right\rangle \right]
+\left( \mu _{+}-\mu _{-}\right) \left\langle \psi ^{\ast }\chi
\right\rangle }. 
\]%
Setting $\mu _{+}=30$ and $\mu _{-}=1$ would correspond to using RMCS as a
unit of time. Using the naive ${\cal E}=2\tanh \left( E/2\right) $ with $E=1$
and computing the appropriate averages from the measured profiles (Fig. 4),
this relation predicts $v\approx 0.004$. Being both of the right sign and
order of magnitude, this is an encouraging result.

There is an alternative approach, based on a {\em discrete} version of a
simple hopping model. Without delving into the details, we only state that
it can be obtained as an approximation from the full master equation. By
considering $\left\langle \sigma \left( \vec{x},t\right) \right\rangle $ and
ignoring all correlations, the result is a discrete equation for the average
densities $\rho _{\pm }\left( \vec{x},t\right) $. Focusing only on the $y$
co-ordinate (i.e., averaging the densities over $x$), our starting point is 
\begin{equation}
\rho _{\pm }\left( y,t+1\right) -\rho _{\pm }\left( y,t\right) =\frac{1}{4}%
{\mu  \choose 1}%
\left\{ J_{in}-J_{out}\right\} ,  \label{start}
\end{equation}%
where the jumps into/out-of site $y$ are given by 
\begin{eqnarray*}
J_{in} &\equiv &\rho _{\pm }\left( y\mp 1,t\right) \phi \left( y,t\right)
+e^{-E}\rho _{\pm }\left( y\pm 1,t\right) \phi \left( y,t\right) \\
J_{out} &\equiv &\rho _{\pm }\left( y,t\right) \phi \left( y\pm 1,t\right)
+e^{-E}\rho _{\pm }\left( y,t\right) \phi \left( y\mp 1,t\right) .
\end{eqnarray*}%
Here the time unit is a RMCS, while the factor $1/4$ accounts for half of
jumps being transverse, so that each term in the $J$'s represents jumps with 
$1/4$ probability. Of course, periodic boundary conditions are imposed: $%
\rho _{\pm }\left( y,t\right) =\rho _{\pm }\left( y+L,t\right) $. For an
inhomogeneous drifting steady state, we seek functions of the form $\rho
_{\pm }\left( y,t\right) =\rho _{\pm }^{\ast }\left( u\right) $, so that we
would write {\em naively} 
\begin{eqnarray*}
\rho _{\pm }^{\ast }\left( u-v\right) -\rho _{\pm }^{\ast }\left( u\right) 
&=&\frac{1}{4}%
{\mu  \choose 1}%
\left\{ \rho _{\pm }^{\ast }\left( u\mp 1\right) \phi ^{\ast }\left(
u\right) -\rho _{\pm }^{\ast }\left( u\right) \phi ^{\ast }\left( u\pm
1\right) \right.  \\
&&+\left. e^{-E}\left[ \rho _{\pm }^{\ast }\left( u\pm 1\right) \phi ^{\ast
}\left( u\right) -\rho _{\pm }^{\ast }\left( u\right) \phi ^{\ast }\left(
u\mp 1\right) \right] \right\} .
\end{eqnarray*}%
Expecting $v$ to be of the order of $10^{-3}$, the first term on the left
must be modified if we wish to apply this equation to data analysis. We
believe a good approximation is to interpolate the left hand side: 
\[
\rho _{\pm }^{\ast }\left( u-v\right) -\rho _{\pm }^{\ast }\left( u\right)
\simeq v\left[ \rho _{\pm }^{\ast }\left( u-1\right) -\rho _{\pm }^{\ast
}\left( u\right) \right] 
\]%
Rearranging the result, we have 
\begin{eqnarray*}
4v\left[ \rho _{\pm }^{\ast
}\left( u-1\right)- \rho _{\pm }^{\ast }\left( u\right) \right] &=& %
{\mu  \choose 1}%
\left\{ e^{-E}\rho _{\pm }^{\ast }\left( u\pm 1\right) \phi ^{\ast }\left(
u\right)-\rho _{\pm }^{\ast }\left( u\right) \phi ^{\ast }\left( u\pm
1\right) \right.  \\
&+& \left. \rho _{\pm }^{\ast }\left( u\mp 1\right) \phi ^{\ast }\left(
u\right) -e^{-E}\rho _{\pm }^{\ast }\left( u\right) \phi ^{\ast }\left( u\mp
1\right) \right\} 
\end{eqnarray*}%
which can be ``integrated'' once, as usual. The constants are just the
steady state currents ($J_{\pm }^{\ast }\equiv \hat{y}\cdot \vec{J}_{\pm }$%
), so that our basic equations read 
\begin{eqnarray}
4J_{+}^{\ast } &=&\mu \left[ \rho _{+}^{\ast }\left( u\right) \phi ^{\ast
}\left( u+1\right) -e^{-E}\rho _{+}^{\ast }\left( u+1\right) \phi ^{\ast
}\left( u\right) \right] -4v\rho _{+}^{\ast }\left( u\right)  \label{J+} \\
4J_{-}^{\ast } &=&\left[ e^{-E}\rho _{-}^{\ast }\left( u\right) \phi ^{\ast
}\left( u+1\right) -\rho _{-}^{\ast }\left( u+1\right) \phi ^{\ast }\left(
u\right) \right] -4v\rho _{-}^{\ast }\left( u\right) \quad . \label{J-} 
\end{eqnarray}%
In principle, these are recursive relations for the profiles and, imposing
constraints ($\sum \rho _{\pm }^{\ast }=N_{\pm }$), the unknown profiles and
constants ($J_{\pm }^{\ast }$ and $v$) can be found. In practice, this
method is not simple. Here, let us simply test these relations against the
profiles from the data. By regarding $e^{-E}$ also as an unknown, we can
write four {\em linear }equations for these constants and see how well the
agreement is. For simplicity, in a manner similar to the continuum study
above, we choose to sum Eqns. (\ref{J+},\ref{J-}) for two equations. For the
other pair, we first divide Eqns. (\ref{J+},\ref{J-}) by $\rho _{\pm }^{\ast
}\left( u\right) $ respectively and then perform the sum. The results are$%
\allowbreak $%
\[
\left[ 
\begin{array}{c}
J_{+}^{\ast } \\ 
J_{-}^{\ast } \\ 
v \\ 
e^{-E}%
\end{array}%
\right] \simeq \left[ 
\begin{array}{c}
0.99\times 10^{-2} \\ 
-0.65\times 10^{-3} \\ 
0.63\times 10^{-3} \\ 
0.39%
\end{array}%
\right] 
\]%
Again, we are encouraged by how well this crude scheme functions, since the
measured $v$ is about $0.9\times 10^{-3}$ and $e^{-1}\simeq 0.37$. As for
the currents, their small values are typical of jammed states. Though our
data for them are too noisy for a meaningful comparison, we are quite
satisfied by the relative magnitudes and signs.

\section{Summary and Outlook}

We have reported simulation results and mean-field arguments for jamming
transitions in a three-state lattice gas under non-equilibrium conditions.
In our simple model, positive and negative particles diffuse on a lattice,
subject to an excluded volume constraint and an external drive which biases
their motion in opposite directions. Generalizing earlier studies of this
model\cite{HSZ}, we focus here on the effect of differential mobility,
allowing one species to be more mobile than the other by a factor of $\mu $.
With equal numbers of the two species in the system, our key findings are as
follows: The previously observed phase transition persists for the range $%
\mu \in \left[ 1,300\right] $, i.e., a free-flowing state, for low densities
($\bar{m}$) and drive ($E$), giving way to a jammed phase at higher $\bar{m}$
and $E$. Moreover, there is little quantitative change to the $\bar{m}$-$E$
phase diagram. Crossing the transition line at low $\bar{m}$ and high $E$,
we observe hysteresis loops in the density structure factor, consistent with
first order transitions. Across the high $\bar{m}$ and low $E$ portion of
the line, the transitions are continuous. While the phase diagram is
essentially unaffected by $\mu >1$, both disordered and ordered phases
exhibit a systematic drift, i.e., non-vanishing mass current. Focusing on
ordered structures, we find that the cluster drifts in the direction of the
more mobile species. According to the data, hole and charge density profiles
depend in subtle ways on $\mu $, and their drift velocity appears to
saturate as $\mu $ increases. To shed more light on these findings, we
present a mean-field theory, in the form of coupled, nonlinear partial
differential equations of motion for the hole and charge densities.
Homogeneous solutions, corresponding to disordered phases, follow trivially
from the conservation laws for the two particle numbers. A linear stability
analysis demonstrates the presence of an instability, as the overall density
increases. Unfortunately, we were not able to find inhomogeneous \emph{%
drifting} solutions, so that we can only offer several consistency checks
between equations and data, in order to build confidence in our theoretical
description.

Many questions remain open for further study. First of all, a better
understanding of the drift velocity would be desirable. Are there
inhomogeneous solutions to our mean-field equations with non-trivial drift?
If a proof of their absence can be established, then we should include
noise terms (to promote the mean-field equations to Langevin equations) and
study the effects of fluctuations and correlations. Perhaps these will
renormalize the ``bare'' co-efficients in Eqn. (\ref{EoM}) so as to admit
drifting structures. Such analysis, along with further simulations, should
also settle the question whether the velocity continues to grow or
approaches a saturation value as $\mu $ increases. In addition, the noise
terms will allow us to investigate particle correlations and structure
factors. As shown in preceding studies \cite{KSZ97b}, these can be extremely
interesting, even in the disordered phase. Going beyond \emph{equal-time}
structure factors, a study of \emph{dynamic} correlations would provide
considerable insight into currents and drifts. Finally, we note that, \emph{%
at }the first order line, we should find \emph{coexisting} phases. Since the
mass current is significant in the disordered region but quite small in the
jammed region, these states will undoubtedly display a rich variety of
behavior.

Nearly all the simulations performed here are based on a $40\times 40$
lattice. For fixed $E$, the observed transitions cannot possibly persist as the
longitudinal size ($L_y$) goes to infinity. At the mean-field level, the
inhomogeneous solutions effectively depend only on the product $\mathcal{E}%
L_y$. Thus, performing simulations with a range of $E$ and $L_y$ would be
useful for establishing the presence of ``$EL_y$ scaling'' or proving its
absence. In general, explorations of finite size effects are clearly crucial
before reliable conclusions on the nature of the phases and transitions can
be drawn.

A natural expansion of parameter space is to allow for unequal numbers of
the species. Such systems exhibit drifting structures even in the \emph{%
absence} of differential mobility\cite{LZ}. As a consequence, one might
wonder whether it is possible to adjust these particle numbers and $\mu $
such that the jam becomes stationary. Viewed as traffic problems, involving
cars and trucks at different densities and mobilities, the answers to these
questions may shed some light on the essence of traffic jams.

\emph{Acknowledgements.}
We thank A. Vasudevan for helpful discussions.  This research was supported in part by 
a grant from the US National Science Foundation through the Division of Materials Research.

\end{document}

%% file: epsf.tex
\ifx\epsfannounce\undefined \def\epsfannounce{\immediate\write16}\fi
 \epsfannounce{This is `epsf.tex' v2.7k <10 July 1997>}%
\newread\epsffilein    
\newif\ifepsfatend     
\newif\ifepsfbbfound   
\newif\ifepsfdraft     
\newif\ifepsffileok    
\newif\ifepsfframe     
\newif\ifepsfshow      
\epsfshowtrue          
\newif\ifepsfshowfilename 
\newif\ifepsfverbose   
\newdimen\epsfframemargin 
\newdimen\epsfframethickness 
\newdimen\epsfrsize    
\newdimen\epsftmp      
\newdimen\epsftsize    
\newdimen\epsfxsize    
\newdimen\epsfysize    
\newdimen\pspoints     
\pspoints = 1bp        
\epsfxsize = 0pt       
\epsfysize = 0pt       
\epsfframemargin = 0pt 
\epsfframethickness = 0.4pt 
\def\epsfbox#1{\global\def\epsfllx{72}\global\def\epsflly{72}%
   \global\def\epsfurx{540}\global\def\epsfury{720}%
   \def\lbracket{[}\def\testit{#1}\ifx\testit\lbracket
   \let\next=\epsfgetlitbb\else\let\next=\epsfnormal\fi\next{#1}}%
%
%
\def\epsfgetlitbb#1#2 #3 #4 #5]#6{%
   \epsfgrab #2 #3 #4 #5 .\\%
   \epsfsetsize
   \epsfstatus{#6}%
   \epsfsetgraph{#6}%
}%
\def\epsfnormal#1{%
    \epsfgetbb{#1}%
    \epsfsetgraph{#1}%
}%
\newhelp\epsfnoopenhelp{The PostScript image file must be findable by
TeX, i.e., somewhere in the TEXINPUTS (or equivalent) path.}%
\def\epsfgetbb#1{%
%
%
    \openin\epsffilein=#1
    \ifeof\epsffilein
        \errhelp = \epsfnoopenhelp
        \errmessage{Could not open file #1, ignoring it}%
    \else                       
        {
            \chardef\other=12
            \def\do##1{\catcode`##1=\other}%
            \dospecials
            \catcode`\ =10
            \epsffileoktrue         
            \epsfatendfalse     
            \loop               
                \read\epsffilein to \epsffileline
                \ifeof\epsffilein 
                \epsffileokfalse 
            \else                
                \expandafter\epsfaux\epsffileline:. \\%
            \fi
            \ifepsffileok
            \repeat
            \ifepsfbbfound
            \else
                \ifepsfverbose
                    \immediate\write16{No BoundingBox comment found in %
                                    file #1; using defaults}%
                \fi
            \fi
        }
        \closein\epsffilein
    \fi                         
    \epsfsetsize                
    \epsfstatus{#1}%
}%
%
\def\epsfclipon{\def\epsfclipstring{ clip}}%
\def\epsfclipoff{\def\epsfclipstring{\ifepsfdraft\space clip\fi}}%
\epsfclipoff 
%
%
\def\epsfspecial#1{%
     \epsftmp=10\epsfxsize
     \divide\epsftmp\pspoints
     \ifnum\epsfrsize=0\relax
       \special{PSfile=\ifepsfdraft psdraft.ps\else#1\fi\space
                llx=\epsfllx\space
                lly=\epsflly\space
                urx=\epsfurx\space
                ury=\epsfury\space
                rwi=\number\epsftmp
                \epsfclipstring
               }%
     \else
       \epsfrsize=10\epsfysize
       \divide\epsfrsize\pspoints
       \special{PSfile=\ifepsfdraft psdraft.ps\else#1\fi\space
                llx=\epsfllx\space
                lly=\epsflly\space
                urx=\epsfurx\space
                ury=\epsfury\space
                rwi=\number\epsftmp
                rhi=\number\epsfrsize
                \epsfclipstring
               }%
     \fi
}%
%
\def\epsfframe#1%
{%
  \leavevmode                   
  \setbox0 = \hbox{#1}%
  \dimen0 = \wd0                                
  \advance \dimen0 by 2\epsfframemargin         
  \advance \dimen0 by 2\epsfframethickness      
  \vbox
  {%
    \hrule height \epsfframethickness depth 0pt
    \hbox to \dimen0
    {%
      \hss
      \vrule width \epsfframethickness
      \kern \epsfframemargin
      \vbox {\kern \epsfframemargin \box0 \kern \epsfframemargin }%
      \kern \epsfframemargin
      \vrule width \epsfframethickness
      \hss
    }
    \hrule height 0pt depth \epsfframethickness
  }
}%
\def\epsfsetgraph#1%
{%
   %
   %
   \leavevmode
   \hbox{
     \ifepsfframe\expandafter\epsfframe\fi
     {\vbox to\epsfysize
     {%
        \ifepsfshow
            \vfil
            \hbox to \epsfxsize{\epsfspecial{#1}\hfil}%
        \else
            \vfil
            \hbox to\epsfxsize{%
               \hss
               \ifepsfshowfilename
               {%
                  \epsfframemargin=3pt 
                  \epsfframe{{\tt #1}}%
               }%
               \fi
               \hss
            }%
            \vfil
        \fi
     }%
   }}%
   %
   %
   \global\epsfxsize=0pt
   \global\epsfysize=0pt
}%
%
%
\def\epsfsetsize
{%
   \epsfrsize=\epsfury\pspoints
   \advance\epsfrsize by-\epsflly\pspoints
   \epsftsize=\epsfurx\pspoints
   \advance\epsftsize by-\epsfllx\pspoints
%
%
   \epsfxsize=\epsfsize{\epsftsize}{\epsfrsize}%
   \ifnum \epsfxsize=0
      \ifnum \epsfysize=0
        \epsfxsize=\epsftsize
        \epsfysize=\epsfrsize
        \epsfrsize=0pt
%
%
      \else
        \epsftmp=\epsftsize \divide\epsftmp\epsfrsize
        \epsfxsize=\epsfysize \multiply\epsfxsize\epsftmp
        \multiply\epsftmp\epsfrsize \advance\epsftsize-\epsftmp
        \epsftmp=\epsfysize
        \loop \advance\epsftsize\epsftsize \divide\epsftmp 2
        \ifnum \epsftmp>0
           \ifnum \epsftsize<\epsfrsize
           \else
              \advance\epsftsize-\epsfrsize \advance\epsfxsize\epsftmp
           \fi
        \repeat
        \epsfrsize=0pt
      \fi
   \else
     \ifnum \epsfysize=0
       \epsftmp=\epsfrsize \divide\epsftmp\epsftsize
       \epsfysize=\epsfxsize \multiply\epsfysize\epsftmp
       \multiply\epsftmp\epsftsize \advance\epsfrsize-\epsftmp
       \epsftmp=\epsfxsize
       \loop \advance\epsfrsize\epsfrsize \divide\epsftmp 2
       \ifnum \epsftmp>0
          \ifnum \epsfrsize<\epsftsize
          \else
             \advance\epsfrsize-\epsftsize \advance\epsfysize\epsftmp
          \fi
       \repeat
       \epsfrsize=0pt
     \else
       \epsfrsize=\epsfysize
     \fi
   \fi
}%
%
%
\def\epsfstatus#1{
   \ifepsfverbose
     \immediate\write16{#1: BoundingBox:
                  llx = \epsfllx\space lly = \epsflly\space
                  urx = \epsfurx\space ury = \epsfury\space}%
     \immediate\write16{#1: scaled width = \the\epsfxsize\space
                  scaled height = \the\epsfysize}%
   \fi
}%
%
%
{\catcode`\%=12 \global\let\epsfpercent=
\global\def\epsfatend{(atend)}%
%
%
%
%
%
%
%
\long\def\epsfaux#1#2:#3\\%
{%
   \def\testit{#2}
   \ifx#1\epsfpercent           
       \ifx\testit\epsfbblit    
            \epsfgrab #3 . . . \\%
            \ifx\epsfllx\epsfatend 
                \global\epsfatendtrue
            \else               
                \ifepsfatend    
                \else           
                    \epsffileokfalse
                \fi
                \global\epsfbbfoundtrue
            \fi
       \fi
   \fi
}%
%
%
\def\epsfempty{}%
\def\epsfgrab #1 #2 #3 #4 #5\\{%
   \global\def\epsfllx{#1}\ifx\epsfllx\epsfempty
      \epsfgrab #2 #3 #4 #5 .\\\else
   \global\def\epsflly{#2}%
   \global\def\epsfurx{#3}\global\def\epsfury{#4}\fi
}%
%
%
\def\epsfsize#1#2{\epsfxsize}%
%
%
\let\epsffile=\epsfbox
 